# Magnetic and electrical properties and carrier doping effects on the Fe-based host compound $Sr_4Sc_2Fe_2As_2O_6$

S. B. Zhang,[1,*] Y. F. Guo,[1] Y. G. Shi,[2,3] S. Yu,[2] J. J. Li,[2,4] X. X. Wang,[2,4]
M. Arai,[5] K. Yamaura,[2,3,4] E. Takayama-Muromachi[3,4]

[1] International Center for Materials Nanoarchitectonics (MANA), National Institute for Materials Science, Tsukuba, Ibaraki 305-0044, Japan

[2] Superconducting Materials Center, National Institute for Materials Science, 1-1 Namiki, Tsukuba, 305-0044 Ibaraki, Japan

[3] JST, Transformative Research-Project on Iron Pnictides (TRIP), Tsukuba, Ibaraki 305-0044, Japan

[4] Department of Chemistry, Graduate School of Science, Hokkaido University, Sapporo, Hokkaido 060-0810, Japan

[5] Computational Materials Science Center, National Institute for Materials Science, 1-1 Namiki, Tsukuba, Ibaraki 305-0044, Japan

Additional charge carriers were introduced to the iron oxyarsenide $Sr_4Sc_2Fe_2As_2O_6$ under a high-pressure condition, followed by measurements of electrical resistivity, Hall coefficient, and magnetic susceptibility. The host compound $Sr_4Sc_2Fe_2As_2O_6$ shows metallic conductivity down to ~200 K and turns to show a semiconducting-like conductivity accompanied by a positive magneto-resistance (22% at 70 kOe). Although the carrier density is comparable at 300 K ($5.9\times10^{21}$ $cm^{-3}$) with that of the other Fe-based superconductors, no superconductivity appears down to 2 K. This is primarily because the net carrier density decreases over 3 orders of magnitude on cooling and additionally a possible magnetic order at ~120 K prevents carriers from pairing. The properties were altered largely by introducing the additional carriers.

## I. INTRODUCTION

Since the discovery of superconductivity (SC) of the Fe based compound [1], various efforts were made to search for new superconductors having much higher $T_c$. The highest $T_c$ reached 57 K to date in the F doped SmFeAsO, which crystallizes into the ZrCuSiAs-type structure [2, 3]. Subsequently, many related Fe based compounds comprising the $Fe_2X_2$ layer (X: pnictogen or chalcogen) and some were found to show SC by chemical or physical doping within first few years. For example, K-doped $BaFe_2As_2$ ($ThCr_2Si_2$-type) shows SC at $T_c$ of 38 K [4-7], $A_xFeAs$ (A = Li and Na) shows $T_c$ of 17 K [8, 9], and FeSe shows $T_c$ of 8 K [10,11], which Tc achieves 37 K under high pressure [12, 13]. It commonly appears that SC emerges in the vicinity of the magnetic ordered state and the structure transition boundary, provoking an open question if SC and the magnetic and structure features are intimately coupled. To answer the question, further studies seem to be required experimentally and theoretically.

Many studies suggested that $T_c$ rises by changing degree of layer to layer correlations [14], as was observed for the cuprate superconductors. $T_c$ of the most intensely correlated layer to layer systems such as $La_{1-x}Sr_xCuO_4$ and $Bi_2Sr_2CuO_6$ is ~38 K in maximum, while 85-92 K for the moderately correlated systems such as $YBa_2Cu_3O_7$ and $Bi_2Sr_2CaCu_2O_8$. Further distant layer to layer systems such as $Bi_2Sr_2Ca_2Cu_3O_{10}$ have $T_c$ of ~123 K in maximum [15], implying a possible way to increase $T_c$ of the Fe-based superconductor.

Taking an overview of the Fe-based compounds, $Sr_4Sc_2Fe_2P_2O_6$ has distant feature of the $Fe_2P_2$ layers and indeed shows the highest $T_c$ of 17 K in the As-free Fe-based oxypnictides [16]. The structure of $Sr_4Sc_2Fe_2P_2O_6$ consists of the perovskite-realted $Sr_2ScO_3$ layers and the $Fe_2P_2$ layers, alternatively stacking up [17, 18]. Although many Fe-based compounds contain the perovskite-related layers, such as $A_3M_2Fe_2(As,P)_2O_5$ [16,19] and $A_4M_2Fe_2(As,P)_2O_6$ [20-23] (A = Ca, Sr, Ba, M = Sc, V, Cr), only 2 members $Sr_4Sc_2Fe_2P_2O_6$ [16] and $Sr_4V_2Fe_2As_2O_6$ [20] are superconducting ($T_c$ is 17 K and 37 K, respectively) as far as we know. It appears that progress of studies of the distant $Fe_2As_2$ layered systems seems to be much behind that of the studies of the intensely coupled $Fe_2As_2$ layered systems such as the doped LnFeAsO (Ln = rare-earth element) and $AFe_2As_2$ (A = Ca, Sr, Ba). We thus focused on a specific Fe-based compound that has increasing distance between the $Fe_2As_2$ layers; we chose $Sr_4Sc_2Fe_2As_2O_6$ [24] because of prospect for appearance of much higher $T_c$ by a chemical doping.

In order to dope additional charge carriers into the $Fe_2As_2$ layers of $Sr_4Sc_2Fe_2As_2O_6$, we designed synthesis to accommodate an amount of oxygen vacancies because a recent theoretical study by a density functional calculation of the local substitute electron density demonstrates that chemical substitutions of such as Co and Ni for Fe do not supply charge carriers [25]. The $d$ electrons of the doped Co and Ni are almost located within the muffin-tin sphere of the substituted atoms. Thus, we believed that oxygen deficiency introduction is a promising way to effect carrier doping into the $Fe_2As_2$ layers, as was successful in the oxyarsenides such as $LaFeAsO_{1-\delta}$ [26] and $TbFeAsO_{1-\delta}$ [27].

In this study, we synthesized the polycrystalline $Sr_4Sc_2Fe_2As_2O_{6-\delta}$ by ambient- and high-pressure synthesis methods and introduced amount of oxygen vacancies to the structure. Effect of the oxygen deficiencies was investigated on the structure, electronic transport, and magnetic properties.

## II. EXPERIMENT

Polycrystalline samples of $Sr_4Sc_2Fe_2As_2O_{6-\delta}$ were prepared by a solid-state reaction in two major steps. First, SrAs powder was prepared from Sr pieces (99.9%, Strem Chemicals) and As grains (99.999%, High Purity Chem.) by heating at 700 ºC for 20 hr in an evacuated quartz tube. Then, SrAs was thoroughly mixed with powders of $Sc_2O_3$ (3N, Furuuchi Chem. Co.), SrO (Lab-made from $SrCO_3$), $Fe_2O_3$ (99.9 %, Furuuchi Chem. Co.) and Fe (99.9%, 100 mesh, Rare Metallic Co.) with proportions at δ of 0, 0.25, 0.5, 0.75, and 1. The mixture each was sealed into a Pt capsule by using a hand-operated press with a BN inner (preheated in nitrogen at 1900 ºC for 1 hr). The inner was installed between the powder and Pt. The capsule was then heated in a belt-type pressure apparatus for 1.5 hr. Elevated pressure and temperature were 3 GPa and 1475 °C, respectively. The capsule was quenched to room temperature before releasing the pressure. Meantime, the oxygen stoichiometric $Sr_4Sc_2Fe_2As_2O_6$ was synthesized under an ambient-pressure condition. The stoichiometric mixture of the powders was heated in a Ta capsule in an evacuated quartz tube at 1050 °C for 50 hr, as reported elsewhere [22]. All of the weighing, grounding, and pressing were processed in a glove-box filled with purified argon except the pressing in the belt-type apparatus.

Chemical phase identification was carried out by an x-ray diffraction (XRD) method in the Rigaku Ultima-IV diffractometer using CuKα radiation. The measurement was achieved at room temperature. Structure analysis by a Rietveld refinement was conducted by using the program Rietica [28]. Electronic resistivity ($\rho$) measurements were carried out in a physical property measurements system (PPMS), Quantum Design, by a four-probe method at a gauge current of 0.5 mA. We measured Hall coefficient ($R_H$) in PPMS by using a horizontal sample rotator in a fixed magnetic field of 50 kOe, eliminating effects of misaligned Hall electrodes. Dc magnetization was measured in a magnetic property measurements system (MPMS), Quantum Design. The sample was cooled without applying a magnetic field ($H$) and then slowly warmed to 300 K in a field of 5 kOe (ZFC; zero-field cooling), followed by cooling down to 2 K again in the field (FC; filed cooling). An isothermal magnetization was measured at 10 K between -10 kOe and 10 kOe in MPMS.

## III. RESULTS AND DISCUSION

The XRD patterns for the samples of $Sr_4Sc_2Fe_2As_2O_{6-\delta}$ are shown in Fig. 1a, indicating that all the samples except the $\delta = 1.0$ sample are of single phase as impurity peaks are significantly small. All observed peaks were well characterized by the tetragonal structure model with the space group *P*4/*nmm* as was reported in Ref. 24, regardless of the high- and ambient-pressure synthesis conditions. The XRD pattern at $\delta = 1.0$ indicates formation of additional phases including $Sr_3As_4$ and Fe. This is probably because the nominal composition is out of range to form $Sr_4Sc_2Fe_2As_2O_{6-\delta}$ at the synthesis condition.

The structure parameters of $Sr_4Sc_2Fe_2As_2O_{5.25}$ were refined by a Rietveld technique as a representative. The refined result is shown in Fig. 2 and the best solution is summarized in Table 1; the analysis was completed successfully with reliable *R* factors. In the analysis, we carefully investigated the oxygen vacancy distribution at the 2 crystallographic site, O1 (4*f* in the Wyckoff position) and O2 (2*c*). The results indicate that the oxygen vacancies are distributed to rather the O2 site than the O1 site.

Please see the crystal structure sketch of $Sr_4Sc_4Fe_2As_2O_6$ (inset to Fig. 2); two perovskite-related layers ($Sr_2ScO_3$; indicated by the solid red lines) are inserted into every interlayer of the $Fe_2As_2$ layers.

If the true oxygen composition is close to the nominal oxygen composition, one oxygen atom in maximum should be absent from the unit cell (indicated by the solid gray lines). The investigation found that the oxygen vacancies distribution is oriented toward the O2 site in the $ScO_2$ layer than the O1 site in the SrO layer.

Eventually, the best solution presented in Table 1 was obtained in the case, supporting the oriented distribution of the oxygen vacancies to the O2 site that should result in supply of electrons into the $Fe_2As_2$ layer. However, it is highly challenging to determine exact quantity of the missing oxygen from the analysis.

By using the structural parameters determined in this study, we estimated the band structure of $Sr_4Sc_2Fe_2As_2O_{6-\delta}$ by means of the first principles calculation; however our preliminary results were not far from what were reported by others [29-31]. We thus decided not to describe details of the electronic structure of the compound.

In Fig.3, nominal oxygen concentration dependence of the lattice parameters $a$, $c$, and the unit cell volume $V$ of $Sr_4Sc_2Fe_2As_2O_{6-\delta}$ is shown. In addition, data for $Sr_4Sc_2Fe_2As_2O_6$ prepared without applying the high-pressure is plotted as well for a comparison. One can see that the lattice parameter $a$ increases with increasing the amount of oxygen deficiencies, while $c$ decreases. $V$ was found to monotonically decrease with increasing the oxygen vacancy concentration. The a-axis expansion may reflect nature of the $Fe_2As_2$ layer, in which the Fe-As bond is slightly stretched out by the electron doping, and the c-axis contraction is mainly owing to the volume reduction of the perovskite block. The anisotropic lattice feature is remarkably contrast with what was found for $LaFeAsO_{1-\delta}$ [26] and $TbFeAsO_{1-\delta}$ [27], in which the unit cell is isotropically contracted with increasing amount of the oxygen vacancies. Nevertheless, the systematic lattice change is indicative of successful doping of the oxygen vacancies into the structure, although we were unable to measure the net oxygen concentration by a chemical method because of the toxic effects.

We found a small gap between $a$ and $c$ of $Sr_4Sc_2Fe_2As_2O_6$ prepared under the high- and ambient-pressure conditions, while $V$ is identical within the experimental accuracy. The gap may reflect a local structure modification caused by the high-pressure heating. In order to reveal the origin of the small gap, additional analysis using a neutron diffraction method for both the samples is needed.

The issue is left for a future work.

However, the results of the refinement of the Rietveld technique does not secure that the nominal quantity of oxygen content is equivalent to the true quantity. So, we tried to check the net oxygen quantity by a gravimetric method. About 20 mg of the samples were full oxidized at 1500 ºC for 24h in air then were slowly cooled to room temperature. The final products should be the oxides of $Sr^{2+}$, $Sc^{3+}$ and $Fe^{3+}$, while As became $As_2O_3$, and evaporated during the sintering process. So, we summarized the residua into one unit formula as $Sr_4Sc_2Fe_2O_{10}$. Then the reaction process could be described by an equation, $Sr_4Sc_2Fe_2As_2O_x + (10-x)/2\ O_2 = Sr_4Sc_2Fe_2O_{10} + As_2O_3\uparrow$. From the weight changes during this process, the net oxygen contents were calculated as following: 5.94, 6.04, 5.83, 5.59, 5.39 and 5.36 for nominal sample δ = 0 (AP), 0 (HP), 0.25 (HP), 0.5 (HP), 0.75 (HP), 1.0 (HP), respectively. The experimental oxygen contents are slightly different to the nominal ones but the agreements are fairly good. It indicates that the oxygen deficiencies have been introduced into samples successfully. And the calculated oxygen contents are close to the nominal ones, considering some errors in the operating process. The experiments were carried two times to ensure the results reliable, and we got two similar results.

Fig. 4 shows temperature dependence of the electrical resistivity of the oxygen-stoichiometric sample of $Sr_4Sc_2Fe_2As_2O_6$ that was prepared under the ambient-pressure condition. The measurement was repeated between 2 K and 300 K in varieties of the applied magnetic field up to 70 kOe. Overall, metallic nature was observed down to 200 K, such as low resistivity ~0.01 Ω-cm at room temperature and positive slope in the plots. An upturn then appeared on cooling at approximately 200 K and a drop at ~125 K, which is associated with a positive magnetoresistivity (MR). Here, the later temperature is defined as $T_{MR}$. The characteristic temperature $T_{MR}$ is almost independent on the strength of the applied magnetic field; $T_{MR}$ is 121.9 K, 122.1 K, 119.9 K, 121.1 K, and 117.9 K for $H$ of 0 kOe, 10 kOe, 30 kOe, 50 kOe, and 70 kOe, respectively. Comparing with the charge transport features of $SrFe_2As_2$ [27], it is suggested that the anomaly is due to formation of the spin density wave (SDW). It is interesting that a comparable anomalous feature of the electrical resistivity, even it is fairly weak, can be seen in the data reported independently by others [24]. To clarify the possibility regarding the formation of the magnetic order, further studies were conducted.

We analyzed the acute increasing of the low temperature resistivity approximately below 70 K. We found the feature is well characterized by the variable rang hopping (VRH) model [32]. The data plot fits well the calculated curve from the analytical equation $\rho = C\exp[(T_0/T)^{1/4}]$, in which C and $T_0$ are constants, as shown in the right inset to Fig. 4. This suggests that the charge carriers are tend to localize on cooling due to inadequate screening of the remote impurity potential at low carrier densities, so that electrons are trapped in random potential fluctuations. Unfortunately, relation to the possible magnetic order and the VRH-type localization remains unclear.

Below $T_{MR}$, an obvious positive MR up to 12.5 % (at 22 K in the field of 70 kOe) was confirmed as shown in Fig. 5, in which $MR = [R(T,H) - R(T,0)]/R(T,0) \times 100$, where $R(T,H)$ and $R(T,0)$ are the resistance with and without $H$, respectively. It is clear that the MR starts to develop on cooling at approximately $T_{MR}$, and it hits the maximum at ~20 K. The temperature at the maximum MR is 21.9 K, 21.1 K, 19.9 K, and 24.1 K in the fields of 10 kOe, 30 kOe, 50 kOe, and 70 kOe, respectively, indicating that it is almost independent on $H$. Inset to Fig. 5 shows the field dependence of MR at different temperature; the MR effect keeps positive and increases with the increasing $H$ at all the temperature points. The MR feature of the compound $Sr_4Sc_2Fe_2As_2O_6$ is distinguishable from that of the related compound $Sr_3Sc_2Fe_2As_2O_5$ [19], which shows positive and negative MR depending on temperature.

MR is usually a powerful tool to investigate nature of electronic scatting in condensed matter [33]. It is possible to attribute the positive MR of $Sr_4Sc_2Fe_2As_2O_6$ to a spin effect caused by a decrease of number of available spin-dependent hopping path, which follows a percolation model for the Mott's VRH model [34, 35]. If the hopping carriers are strongly tied to local spins via ferromagnetic exchange interactions, the Zeeman splitting of the hopping carriers can be comparable to the thermal energy. In the case, the Zeeman splitting can be responsible for the positive MR if a certain fraction of localized states is occupied by two electrons. To describe this in more details, we consider two types of site that contribute to conduction under presence of the on-site Coulomb energy $U$ (assumed to be smaller than the bandwidth of the localized states); a site (A-type) that energy is close to $E_F$ and either unoccupied or singly occupied, and the other site (B-type) that has an occupied deeper level located at $E_F - U$ and accommodates an additional electron having the opposite spin. Hopping between the A and B sites, singly occupied, requires a pair of sites with antiparallel spin, as does the reverse process involving a doubly occupied B site and an unoccupied A site.  The probability of finding pairs of the site with anti-parallel spins decreases with increasing the magnetic field. Although this qualitatively provides an explanation for the observed MR of $Sr_4Sc_2Fe_2As_2O_6$ to some extent, further clarification by experimental and theoretical studies is required.

Fig. 6 shows temperature dependence of $C_p$ of the polycrystalline compound of $Sr_4Sc_2Fe_2As_2O_6$ prepared under the ambient pressure condition.   First, we analyzed the low temperature part of the $C_p$ data; the inset to Fig. 6 shows $C_p/T$ vs. $T^2$ plot of the data below 10 K.   Using the approximate Debye model $C_p(T)/T = \gamma + \beta T^2$, where $\beta$ is a coefficient and $\gamma$ is the Sommerfeld coefficient, we obtained $\beta$ of 1.094 mJ mol$^{-1}$ K$^{-4}$ and $\gamma$ of 10.4 mJ mol$^{-1}$ K$^{-2}$.   By using the relation to the Debye temperature $\Theta_D = (12\pi^4 nR/5\beta)^{1/3}$, where R is the gas constant and $n$ is the number of atoms in the mole, we estimated $\Theta_D$ to be 305 K.   It is noteworthy that $C_p/T$ approaches to 35 mJ mol$^{-1}$ K$^{-2}$ by extrapolation at the low temperature limit.   The $\gamma$ enhancement is possibly owing to magnetic contributions suggested in the MR study.

Next, we tried to find a corresponding anomaly regarding the possible magnetic order suggested at $T_{MR}$.   In order to separate the lattice contribution from the total $C_p$, the data were quantitatively analyzed by using a linear combination of the Debye model, the Einstein model, and the electronic specific heat.   The analytical formula was

$$C(T) = n_D \times 9 N_A k_B \left(\frac{T}{T_D}\right)^3 \int_0^{T_D/T} \frac{x^4 e^x}{(e^x-1)^2} dx + n_E \times 3 N_A k_B \left(\frac{T_E}{T}\right)^2 \frac{e^{T_E/T}}{\left(e^{T_E/T}-1\right)^2} + \gamma T$$,

where $N_A$ is the Avogadro's constant, $k_B$ is the Boltzmann's constant, and $T_E$ is the Einstein temperature.    In the analysis $\gamma$ was fixed to the observed value (10.4 mJ mol$^{-1}$ K$^{-2}$).   The scale factors $n_D$ and $n_E$ correspond to the numbers of vibrating modes per the formula unit in the Debye and the Einstein models, respectively.   Fit to the data yielded $T_D$ of 261(3) K, $T_E$ of 561(10) K, $n_D$ of 9.8(2), and $n_E$ of 6.3(1).   The fit to the data is shown in Fig. 6; the function apparently fits the data well.   $C_p$ however changes smoothly at $T_{MR}$ and thus we could not construct validity of the possible magnetic contribution.   The possible magnetic order actually remains to be studied by additional methods such as NMR and neutron diffraction.

We studied Hall-effect of $Sr_4Sc_2Fe_2As_2O_6$. The measurements were conducted on the ambient-pressure sample. The data are presented in Fig. 7. It appears that $R_H$ is negative over the entire temperature range, indicating that the dominating carriers are electrons, as was found in the parent compound $SrFe_2As_2$ [36-37]. The carrier density at room temperature is $5.9\times10^{21}$ cm$^{-3}$, corresponds to 1.5 electrons per unit cell. The density is comparable with the optimally carrier doped Fe-based superconductor such as $LaFeAsO_{0.85}$ [26] and $SrFe_{1.88}Ni_{0.12}As_2$ [38] (see the plots in Fig. 7) and $LaFeAsO_{0.9}F_{0.1-\delta}$ [39]. The carrier density monotonically decreases on cooling over 3 orders of the magnitude to $9.9\times10^{17}$ cm$^{-3}$ (5 K), corresponds to 0.0003 electrons per unit cell. Although the strong temperature dependence is usually considered due to such as strong multiband effects and spin related scattering effects [40, 41], the appearance of the notable MR effect (see inset to Fig. 5) makes it much likely to be a spin related scattering effect owing to the formation of the possible magnetic order. Comparisons of the data between the SDW and SC compounds such as $SrFe_2As_2$ (SDW at ~200 K [37]) and $SrFe_{1.88}Ni_{0.12}As_2$ ($T_c$ = 9 K [38]); $LaFeAsO$ (SDW at ~160 K [37]) and $LaFeAsO_{0.85}$ ($T_c$ = 26 K [26]) imply the spin related scattering effects at ~120 K for $Sr_4Sc_2Fe_2As_2O_6$. The data of isostructural compound $Sr_4V_2Fe_2As_2O_6$ ($T_c$ = 37 K [20]) also implies the same possibility. Besides, it should be noticed that the $R_H$ and the Hall mobility ($\mu_H$) plots (inset to Fig. 7) each shows an abrupt change near $T_{MR}$, suggesting the spin related scattering effects. If the increasing suppression of $n$ at low temperature results from formation of a magnetic order at $T_{MR}$ such as SDW, the presence of the magnetic order could be a reason of absence of SC in $Sr_4Sc_2Fe_2As_2O_6$. It should be noted that we tried to measure $R_H$ of the additionally carrier doped $Sr_4Sc_2Fe_2As_2O_{6-\delta}$; however the attempt was so-far unsuccessful because of the chemical instability of the doped samples.

Fig. 8a shows temperature dependence of the electrical resistivity of the polycrystalline samples of $Sr_4Sc_2Fe_2As_2O_6$ prepared under the high-pressure (HP) condition. Any traces of metallic features are not visible, unlike the data for the ambient-pressure sample (Fig. 4). The room temperature resistivity is approximately one order of magnetite higher than that of the ambient-pressure sample and the temperature dependence is just like what is expected for a semiconductor. MR effect is not obvious over the whole temperature range at 70 kOe. The rather poor conductivity possibly reflects the local structure change such as degree of distortion/tilting forced by the heating under the high-pressure condition. The XRD study indeed indicated that the lattice parameters $a$ and $c$ varies from the values of the ambient pressure sample without change of the unit-cell volume. The change may alter the electronic state somewhat toward reducing the conductivity. Additionally, resistive grain boundary and others, which are undetected in the XRD study, are possibly formed in the high-pressure heating, masking the true conducting features; the issue is left for future study.

Figs. 8b-8d show temperature and oxygen quantity dependence of the electrical resistivity of $Sr_4Sc_2Fe_2As_2O_{6-\delta}$ prepared under the high-pressure condition. Overall, the electrical conductivity is somewhat ameliorated by introducing the oxygen deficiency. $Sr_4Sc_2Fe_2As_2O_{5.5}$ shows an upturn at approximately 50 K, while $Sr_4Sc_2Fe_2As_2O_{5.25}$ does show metallic behaviors down to 2 K. As we expected, it appears that introduction of the oxygen deficiency makes the perovskite-like block supply electron carriers to the $Fe_2As_2$ layer. However, SC does not appear in any of the samples studied so-far, unfortunately. As discussed in Ref. 23, it is possible that SC of the doped Fe pnictides is provoked not only by increase of the density of state at the Fermi level but also an electronic structural modulation. If this is valid here, the carrier doping would not be enough to tune the electronic state of $Sr_4Sc_2Fe_2As_2O_6$ toward appearance of SC.

Temperature dependences of the magnetic susceptibility of all the samples are shown in Figs. 9a-9e. The measurements were conducted in a magnetic field of 5 kOe. The ambient pressure sample $Sr_4Sc_2Fe_2As_2O_6$ shows paramagnetic behavior without anomaly corresponds to the possible magnetic ordering. Besides, a ferromagnetic contribution (approximately 0.15 $\mu_B$ per the formula unit at 10 K) was detected in the *M-H* measurements (see inset to Fig. 9a). The ferromagnetic part is likely due to a small amount of magnetic impurities, and it is possible that the expected anomaly corresponds to the possible magnetic ordering is masked by the ferromagnetic background. Meanwhile, the high-pressure sample $Sr_4Sc_2Fe_2As_2O_6$ shows much smaller ferromagnetic contribution approximately one thirty, and a kink is somehow visible which is possibly coupled to the magnetic ordering. In order to further study magnetic nature of the carrier doped $Sr_4Sc_2Fe_2As_2O_{6-\delta}$, employing other microscopic techniques such as NMR and μSR would be helpful.

## IV. CONCLUSIONS

The host compound $Sr_4Sc_2Fe_2As_2O_6$ has comparable amount of charge carriers at room temperature when compared to values of the Fe-base superconductors [20, 26, 38]. The compound, however, quickly loses the carriers on cooling over 3 orders of magnitude. At low temperature, only few number of electrons, 0.0048 electrons per unit cell (even at 50 K), is available to transport. This sharply contrasts to features of the Fe-based superconductors such as $LaFeAsO_{0.85}$ ($T_c$ = 26 K) [26] and $Sr_4V_2Fe_2As_2O_6$ ($T_c$ = 37 K) [20] and the Ni doped $SrFe_2As_2$ ($T_c$ = 9 K) [38]; the carrier density of $Sr_4Sc_2Fe_2As_2O_6$ is more than 1-2 orders of magnitude smaller than that of the superconductors even at 50 K (see Fig. 7). The fast decrease of the carrier density on cooling primarily accounts for absence of SC of $Sr_4Sc_2Fe_2As_2O_6$; namely the carrier density at low temperature is too small to establish SC, although the carrier density at room temperature is comparable with that of other Fe-based superconductors. Hence, control of decreasing rate of the carrier density on cooling is rather crucial than adding carriers to establish SC in $Sr_4Sc_2Fe_2As_2O_6$.

The quick loosing of carriers of $Sr_4Sc_2Fe_2As_2O_6$ by cooling possibly results from a delicate balance of negatively and positively charged carriers, those are compensated each other. As predicted by the first principle studies on $Sr_4Sc_2Fe_2As_2O_6$ [29-31], nature of the Fe $3d$ multibands produce the 2 types of carriers simultaneously. The band each may have different electrons-scattering rate with temperature, and it is thus reasonable to see a complex temperature dependence of the Hall coefficient because all of the rates are combined. The balance regarding the electrons-scattering rates may be varied somewhat by replacing Sc for V, resulting in enhancement of temperature dependence of the carrier density. If this is true, we may expect a chemical substitution that does not enhance the temperature dependence and hence helps to establish SC. According to a prediction based on the first principles study on $Sr_4Sc_2Fe_2As_2O_6$, Ti substitution for Sc may be effective for the purpose [29]; further studies of varieties of substitution for the compound are in progress.

In addition, establishment of a magnetic order was suggested for $Sr_4Sc_2Fe_2As_2O_6$ at around 125 K by the MR studies. If this is true, the conducting carriers tend to localize below the temperature and can be additionally responsible for the quick losing of the net carriers. The Hall mobility and Hall coefficient indeed change the characters at $T_{MR}$ on cooling. Regardless of the possibility, we successfully synthesized the additionally carrier doped compounds $Sr_4Sc_2Fe_2As_2O_{6-\delta}$ ($0 \leq \delta < 1$) under the high pressure condition. Although we were unable to measure net oxygen quantity, a certain amount of oxygen vacancies were introduced to some extent, resulting in additional electrons doping to the $Fe_2As_2$ layer. Unfortunately, no SC appears down to 2 K, even though the compound shows metallic nature to the lowest limit. The result sharply contrasts with phenomenology of the superconductor $Sr_4V_2Fe_2As_2O_6$ ($T_c$ = 37 K) [20]. Because of prospect for appearance of higher $T_c$ SC of the doped $Sr_4Sc_2Fe_2As_2O_6$, we would keep studying on the compound.

**ACKNOWLEDGMENTS**

We thank Dr. M. Miyakawa for the high-pressure studies. This research was supported in part by the World Premier International Research Center (WPI) Initiative on Materials Nanoarchitectonics from MEXT, Japan; the Grants-in-Aid for Scientific Research (20360012, 22246083) from JSPS, Japan; and the Funding Program for World-Leading Innovative R&D on Science and Technology (FIRST Program) from JSPS.

**REFERENCES**

*Corresponding author. Tel: 081-029-851-3354-8024; Fax: 081-029-860-4674

E-mail address: ZHANG.Shoubao@nims.go.jp (S. B. Zhang)

**Table I** Structure and isotropic displacement parameters of $Sr_4Sc_2Fe_2As_2O_{5.25}$ prepared under the high pressure condition.[a]

| Atom | Wyckoff position | Site occupation | $x$ | $y$ | $z$ | $B_{iso}(Å^2)$ |
|---|---|---|---|---|---|---|
| Sr1 | 2c | 1 | 0.75 | 0.75 | 0.1862(2) | 0.58(2) |
| Sr2 | 2c | 1 | 0.75 | 0.75 | 0.4147(2) | 0.36(3) |
| Sc  | 2c | 1 | 0.25 | 0.25 | 0.3064(5) | 0.07(0) |
| Fe  | 2a | 1 | 0.25 | 0.75 | 0        | 1.49(1) |
| As  | 2c | 1 | 0.25 | 0.25 | 0.0856(4) | 0.54(2) |
| O1  | 4f | 0.96(1) | 0.25 | 0.75 | 0.2884(3) | 0.32(5) |
| O2  | 2c | 0.80(1) | 0.25 | 0.25 | 0.4300(1) | 4.42(4) |
| Lattice parameters | | | $a$ (Å) | | $b$ (Å) | $c$ (Å) |
| | | | 4.0526(2) | | 4.0526 | 15.7304(3) |
| Refine parameters | | | $R_p$ | | $R_{wp}$ | $\chi^2$ |
| | | | 7.557 | | 9.786 | 9.372 |

[a] Note: Formula sum: Sr4 Sc2 Fe2 As2 O5.44, Formula weight: 788.966 g/mol, Crystal system: tetragonal, Space group $P4/nmm$ (no. 129), $Z = 1$, Cell ratio: $a/b = 1$ and $b/c = 0.2576$ and $c/a = 3.8816$, Cell volume: 258.35 Å$^3$, and Calculated density: 5.0708 g/cm$^3$.

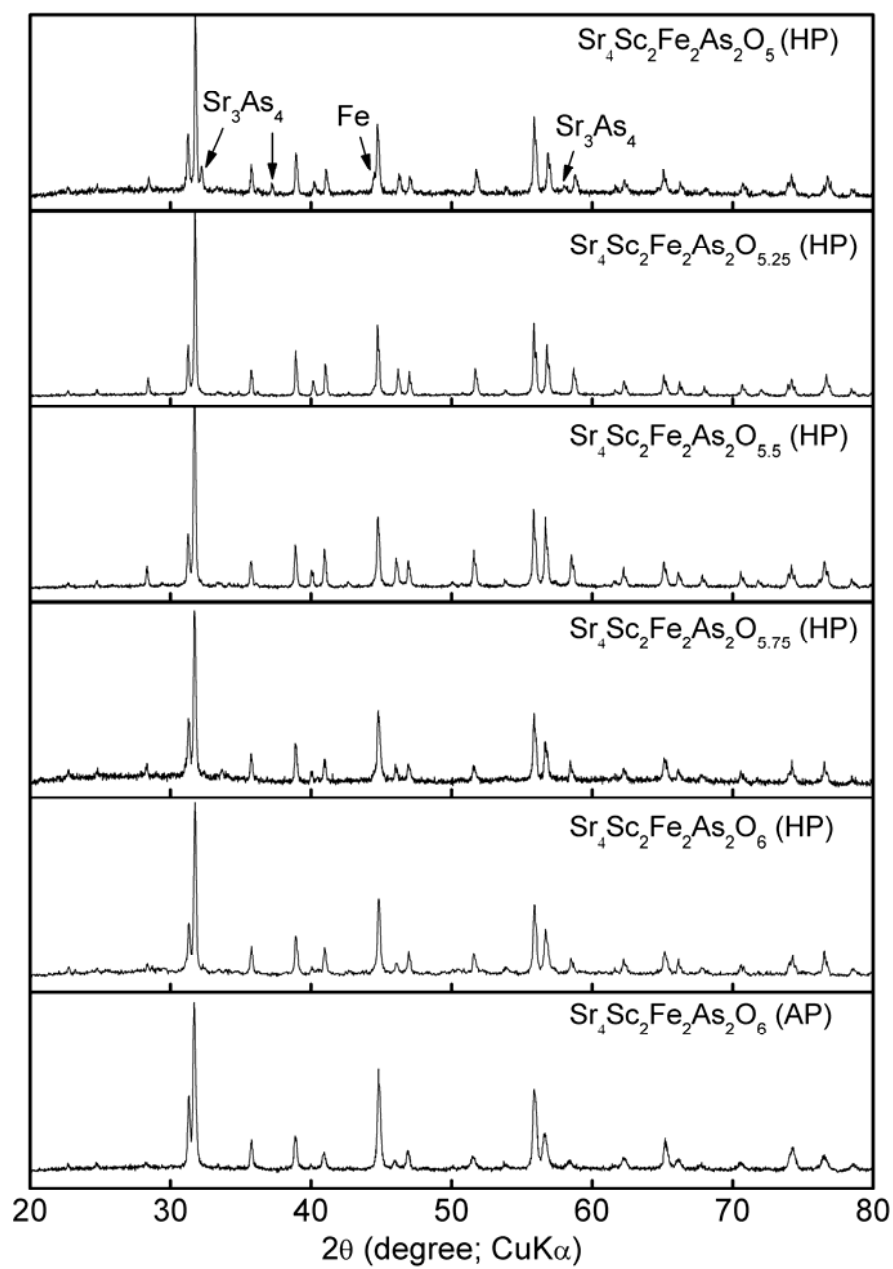

Fig. 1 Powder XRD patterns of the polycrystalline samples of $Sr_4Sc_2Fe_2As_2O_{6-\delta}$ prepared under the high pressure (HP) and ambient pressure (AP) conditions.

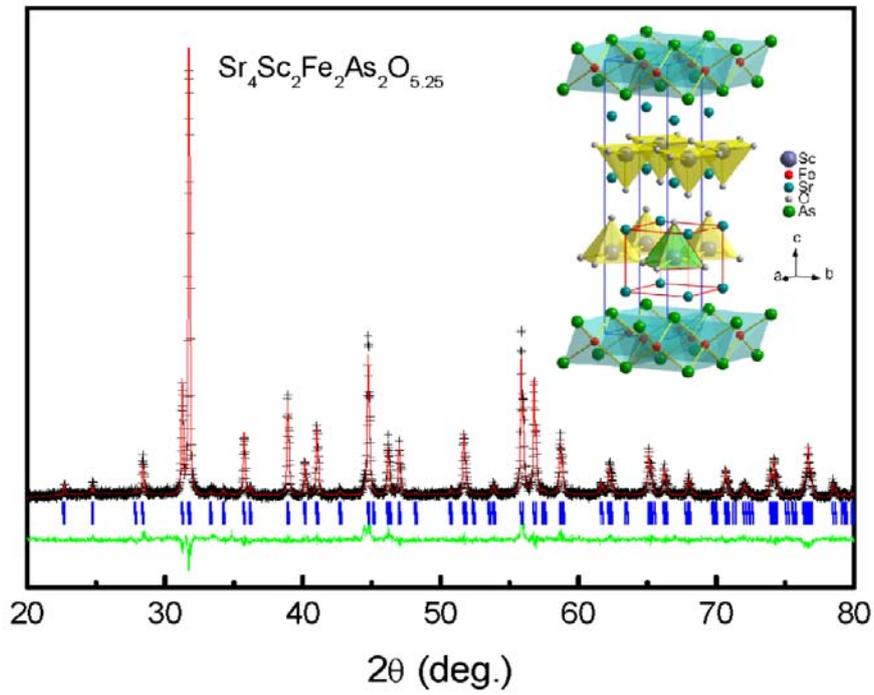

Fig. 2 Rietveld refinement of the powder XRD pattern of $Sr_4Sc_2Fe_2As_2O_{5.25}$: small crosses and solid lines indicate the experimental and calculated data, respectively. The lowest curve corresponds to the difference between the curves. The vertical bars indicate expected Bragg reflection positions.

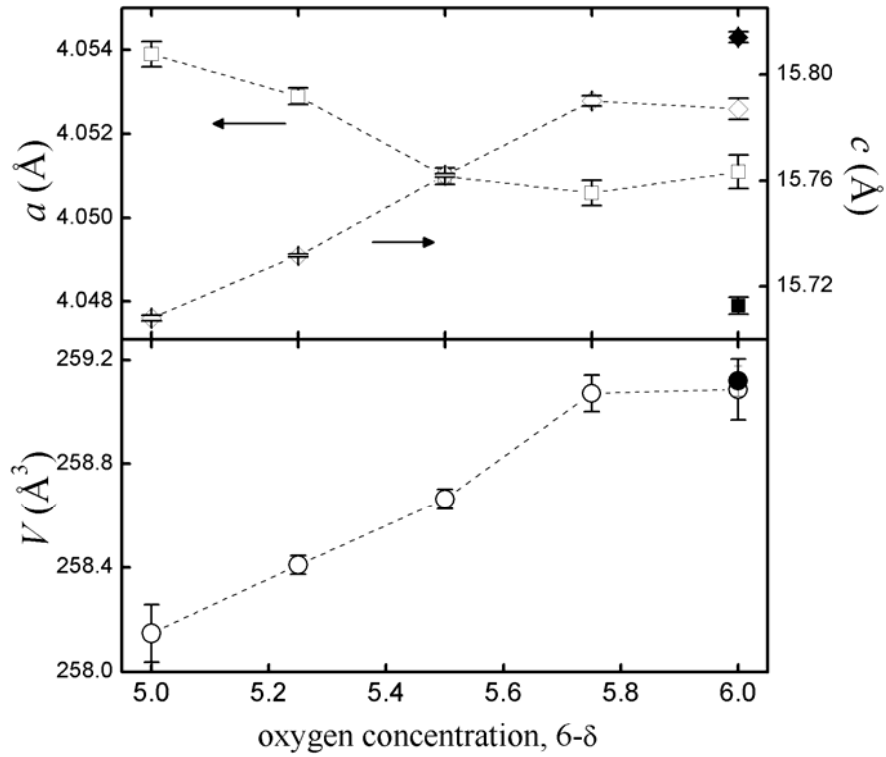

Fig. 3　Lattice parameter variation of $Sr_4Sc_2Fe_2As_2O_{6-\delta}$ determined by the XRD study. Open and closed symbols respectively correspond to the high-pressure and ambient-pressure samples.

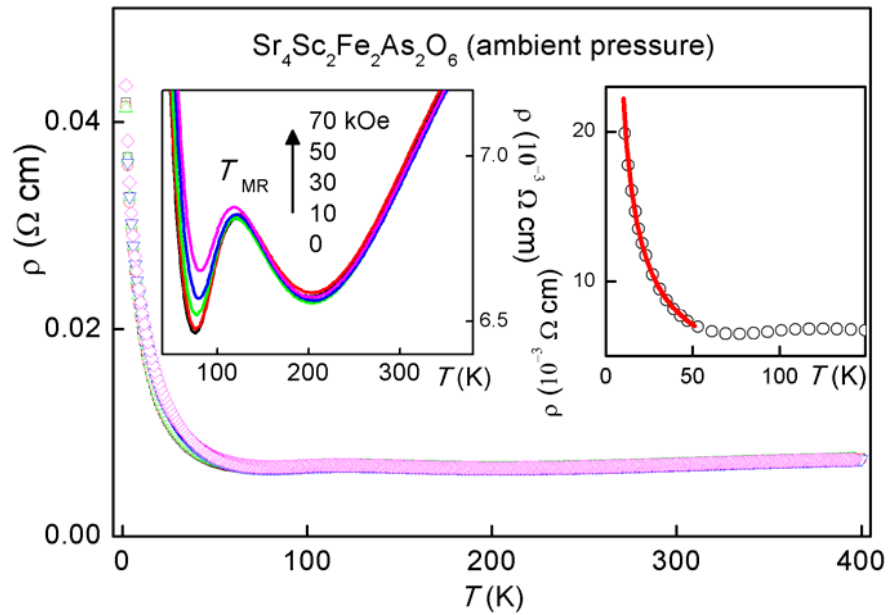

Fig. 4　$T$ dependence of $\rho$ of $Sr_4Sc_2Fe_2As_2O_6$ (ambient pressure) measured in a variety of $H$ (indicated). Insets are an expanded view and a VRH fit (solid curve) to the $H = 0$ data.

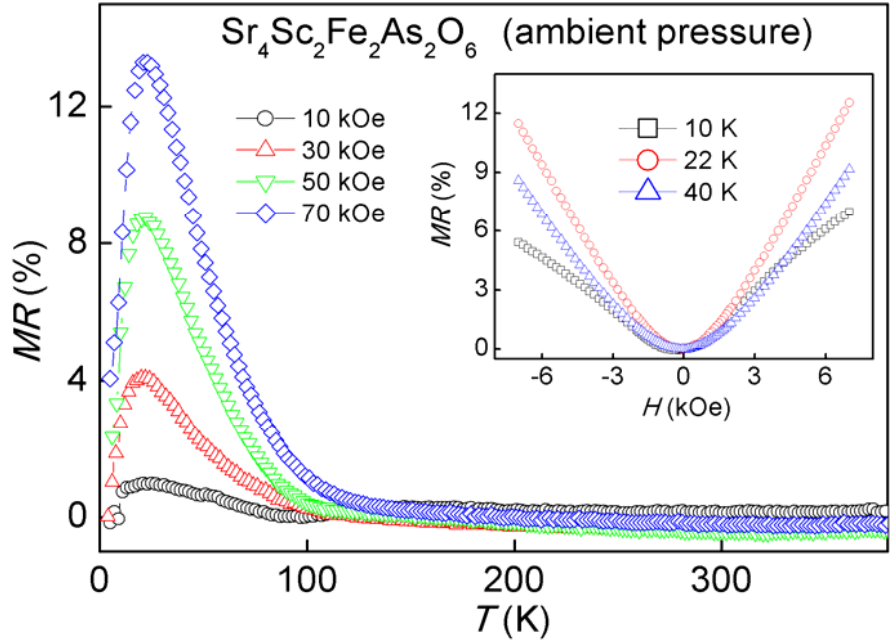

Fig. 5 *T* dependence of *MR* of $Sr_4Sc_2Fe_2As_2O_6$ (ambient pressure) measured in a variety of $\mu_0H$ (indicated).   The inset shows *H* dependence of *MR* at *T* = 10 K, 22 K, and 40 K.

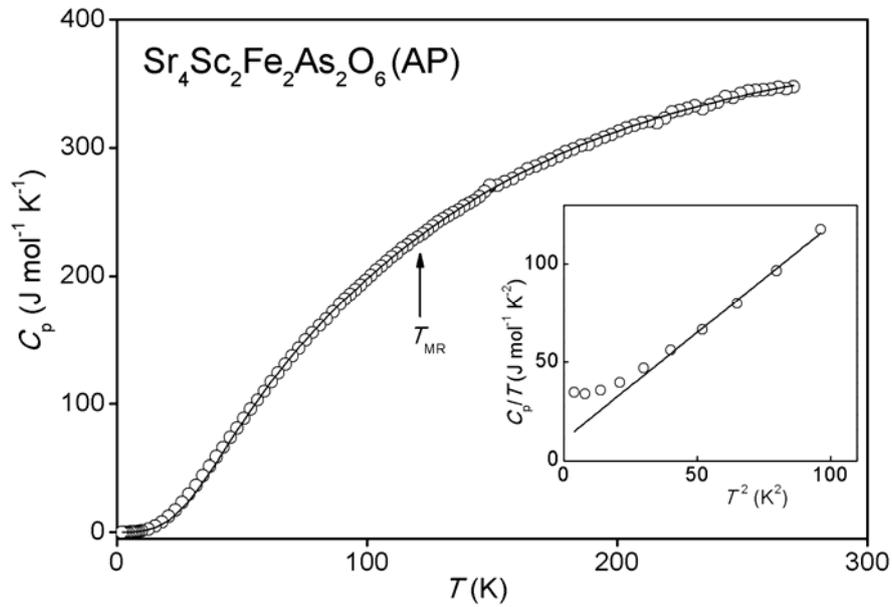

Fig. 6 *T* dependence of $C_p$ of $Sr_4Sc_2Fe_2As_2O_6$ prepared under the ambient-pressure (AP) condition. Inset shows $C_p/T$ vs. $T^2$ plot of the data.   Curve and line are fits to the data.

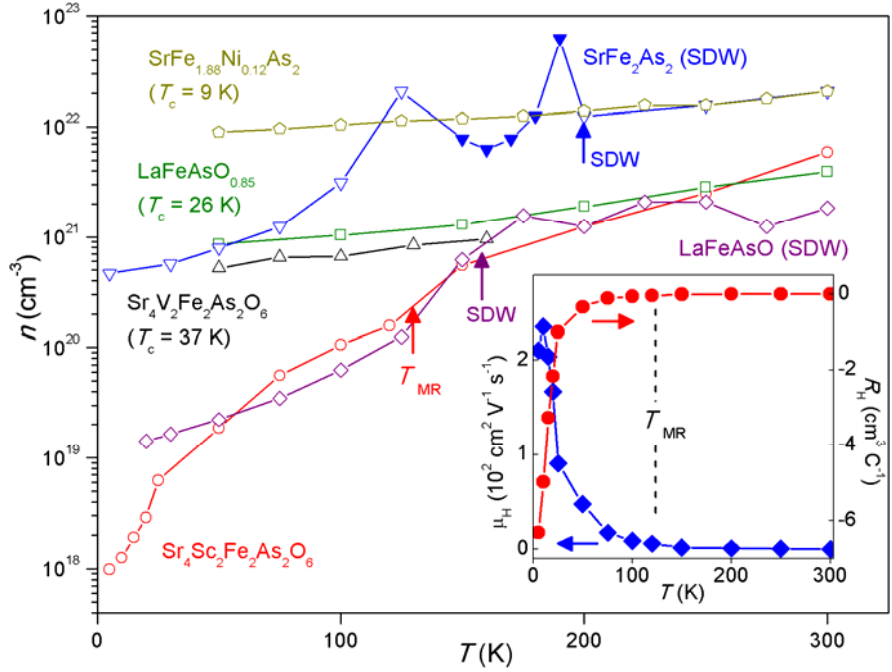

Fig. 7 $T$ dependence of $n$ and $R_H$ (inset) and $\mu_H$ (inset) of the ambient pressure sample $Sr_4Sc_2Fe_2As_2O_6$. The data for $LaFeAsO_{0.85}$, $Sr_4V_2Fe_2As_2O_6$, and the Ni doped $SrFe_2As_2$ were taken from Refs. 24, 18, and 36, respectively. The data for $LaFeAsO$ and $SrFe_2As_2$ were taken from Ref. 35. The solid triangle symbol indicates data calculated from positive $R_H$ values.

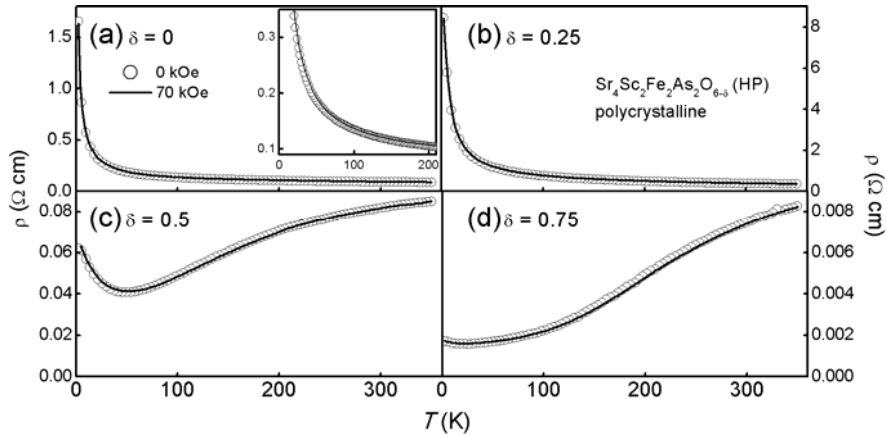

Fig. 8 $T$ dependence of $\rho$ of $Sr_4Sc_2Fe_2As_2O_{6-\delta}$ with and without applying $H$ of 70 kOe. Inset is an expanded view.

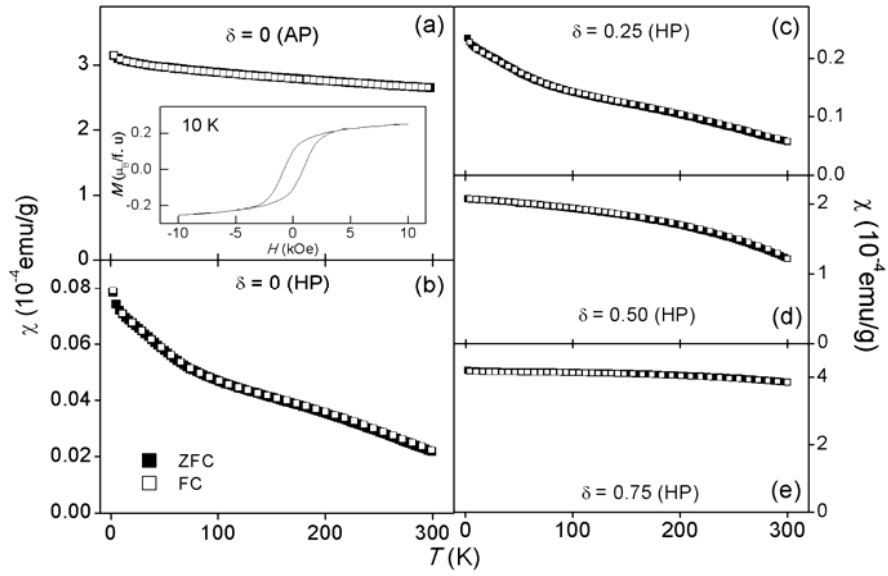

Fig. 9 $T$ dependence of $\chi$ of $Sr_4Sc_2Fe_2As_2O_{6-\delta}$ measured at 5 kOe. Inset is the magnetic loop for $Sr_4Sc_2Fe_2As_2O_6$ (AP) measured at 10 K.